\documentclass[aps,pra,twocolumn,showpacs]{revtex4}
\usepackage{graphicx}%
\usepackage{amsmath}%
\usepackage{amsfonts}%
\usepackage{amssymb}

\begin{document}

\title{Spatial solitons in periodic nanostructures}

\author{A.V. Gorbach and D.V. Skryabin}
\affiliation{Centre for Photonics and Photonic Materials,
Department of Physics, University of Bath, Bath BA2 7AY, UK}

\begin{abstract}
We present the first principle theory of the existence and stability
of TE and TM spatial solitons  in a  subwavelength periodic
semiconductor-dielectric structure. We have found that for the
wavelength $1550$nm and the  interface separation close to and less
than $100$nm the band structure of the linear TE and TM waves
becomes similar to the band structure of a homogeneous medium. The
properties of TE solitons change accordingly. However, the
transverse profiles of the TM solitons continue to reflect the
subwavelength geometry of the structure  and develop dominant
intensity peaks inside the low index dielectric slots. Our stability
analysis based on the linearized Maxwell equations indicates that
the nonlinear TM waves can be approximated as the evanescently
coupled modes of the  slot waveguides with the low index dielectric
core and the high index semiconductor cladding. Transition to the
regime where the slot waveguides start to determine properties of TM
waves is associated with the so called Brewster condition.
\end{abstract}

\pacs{42.65.Tg,78.67.Pt,42.70.Nq}

\maketitle

\section{Introduction}
Recent progress in the fabrication of nanostructures for photonics
applications has stimulated research into light trapping and guiding
on the subwavelength scale \cite{SLY+2005,Maier2006,BFL+2007}.
Surface plasmon polaritons tightly confined to the metal-dielectric
interfaces have been at the focus of recent efforts in this
direction \cite{BDE2003,BVD+2006,EGB2008,RRA+2008}. The research
into plasmons has included studies of their interaction with
periodically structured metals, see, e.g. \cite{WZ2008}, which have
been recently extended to soliton structures in periodic arrays of
metalic slot waveguides filled with a nonlinear dielectric
\cite{LBG+2007}. The latter work has been a significant and
conceptual departure from research into optical solitons in
nonlinear waveguides arrays coupled by evanescent waves
\cite{LSC+2008}. The metal dielectric interfaces in \cite{LBG+2007}
are separated by distances much smaller than the wavelength, so that
the concept of the coupling induced by the evanescent waves
\cite{LSC+2008,kivshar_book}  becomes largely irrelevant. New
effects can be expected in this regime, which are still waiting to
be explored.

Semiconductors also can be used for subwavelength guiding.
In particular, silicon photonic wires have
been recently promoted as promising and close to practical
applications building blocks of photonic chips, where nonlinear and
soliton effects have been already extensively researched \cite{RNM+2007}.
The large refractive index of silicon ($n\simeq 3.5$) allows for tight
light confinement by the conventional total internal reflection
mechanism, giving the simultaneous advantages of strong ultrafast Kerr
nonlinearity ($n_2\simeq 4\times 10^{-18}$m$^2$/W), controlled
dispersion and manageable losses. Losses are a particular
problem for plasmons, suggesting that their nonlinear functionality
is likely to become  more viable  if gain is introduced into
the dielectric \cite{Nez_Opt_Expr,Nog_Opt_Expr}. Two photon and free career induced absorptions
are traditionally thought of as
hampering the attractiveness of silicon
for nonlinear applications, but these often can be lived with
\cite{Ding2008,bullets}
or managed, e.g. by the electrically removing free carriers from the
waveguide core \cite{JRL+2005,RJL+2005,RNM+2007}. There are also other highly nonlinear
semiconductors and  glasses which can be useful  for various on-chip
applications. In particular, many of the soliton experiments in
planar waveguide arrays have been performed using doped GaAs
($n=3.47$, $n_2=3.3\times 10^{-17}$m$^2$/W) structures \cite{Aitchison}.

While an isolated silicon or GaAs photonic wire  confines light
within an area of the order of the wavelength inside the material squared
($\lambda_{vac}^2/n^2$), bringing two wires together with a separation of few tens
of nanometers produces a strong intensity peak {\em in-between the wires}, with the field
predominantly polarised perpendicular to the interface between the
wires (TM-modes). This type of waveguides is  called  slot waveguides
and they provide an elegant method  of focusing light into a subwavelength region
\cite{AXB+2004}. One can ask a question about the existence of solitary waves
having subwavelength dimensions  in  arrays of the
semiconductor-dielectric slot waveguides. The  losses in such arrays
are expected to be few dB/cm \cite{BHW+2005,BSG+2007,SDF+2007}.
Measurements of
the transverse profile of the slot mode have been reported in
\cite{FLC+2008}, while applications of the silicon slot waveguides have been demonstrated for
frequency conversion and wavelength division multiplexing \cite{KVD+2008},
as well as for the design of high Q resonators \cite{BSG+2007} and
sensors based upon them \cite{RCL2008}.

In this work, we consider an infinite
array of semiconductor photonic wires
with the wire widths and separations taken well below the wavelength.
 We assume that the photon energy is below the bandgap and
the semiconductor acts essentially as a high index nonlinear
dielectric. We present analysis of linear and soliton solutions for both
TE and TM polarizations using first principle Maxwell equations.
The TE solitons undergo a smooth transformation from
the case, which can be well understood in terms  of the evanescently coupled waveguides
into the solitons of a quasi-homogeneous medium, when the widths
of all layers  become much less than the wavelength. Conversely and centrally to our work, the
TM solitons, under the same change of geometry, evolve into
structures with the dominant intensity peaks located outside the
semiconductor, i.e. in the low index material. These peaks become the prevailing features as the
separation of high index semiconductor waveguides is reduced.
Thus the TM solitons are not
transformed into the solitons of the quasi-homogeneous medium, which does not distinguish
between polarizations. Our analysis strongly suggests that
for subwavelength separations, the TM solitons can be qualitatively considered as
discrete solitons composed of the coupled modes of the slot
waveguides. We also propose and apply a technique to study linear stability of solitons within
the framework of the linearised Maxwell equations.

\section{Maxwell equations and soliton equations}
\begin{figure}
\includegraphics[width=0.22\textwidth]{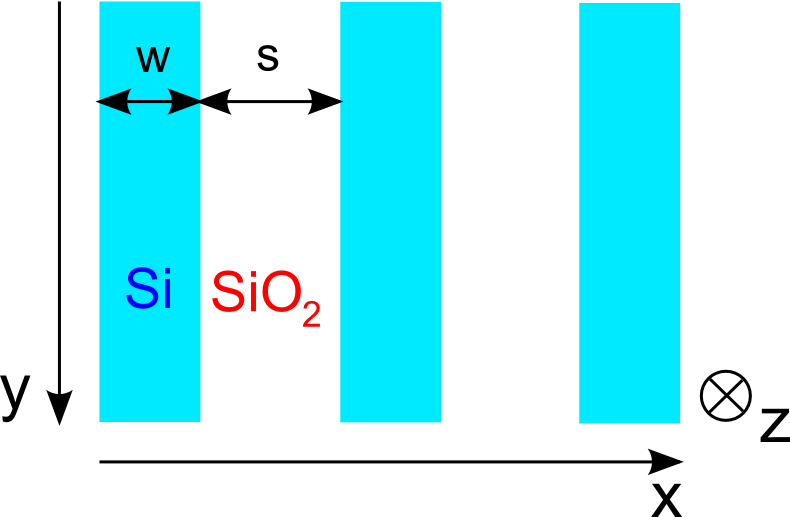}
\caption{(Color online) One dimensional periodic structure consisting of layers of high
index semiconductor such as silicon (width $w$) and low-index dielectric such as silica glass (width $s$).}
\label{fig1}
\end{figure}

The array of slot waveguides we consider below is a periodic
structure of narrow  layers of a high index semiconductor material
(material $s$, e.g., silicon) embedded into a low index dielectric
material (material $g$, e.g., silica glass), see Ref.
\cite{SDF+2007} and Fig. \ref{fig1}. The separations of the
semiconductor layers vary from $500$ to $50$nm, and the width of the
layers is between $220$ and $95$nm. The vacuum wavelength is taken
to be $\lambda_{vac}=1.55\mu$m. Our analysis is based on the
nonlinear Maxwell equations in the 2 dimensional geometry:
\begin{equation}
\vec\nabla\times\vec H=-ikc\epsilon_0\vec D,~~\vec\nabla\times\vec E=i\frac{k}{c\epsilon_0}\vec H, \label{max}
\end{equation}
where $k=2\pi/\lambda_{vac}$, $c$ is the speed of light in vacuum, $\epsilon_0$ is the vacuum
permittivity, and for electric $\vec{\mathcal{E}}$ and magnetic $\vec{\mathcal{H}}$
fields  it is assumed that
$\vec{\mathcal{E}},\vec{\mathcal{H}}=
{1\over 2}\vec{E},\vec{H}\cdot \exp(-ikc t)+c.c.$
Light propagation is assumed to be along the $z$-direction and the $x$-coordinate is perpendicular to the layers,
see Fig.~\ref{fig1}.

For the TM-polarized modes (such that
only $H_y,E_x$ and $E_z$ are non-zero) the Maxwell equations are reduced to
\begin{eqnarray}
\label{tm1v2}
&& \partial_{zz} E_x - \partial_{zx} E_z=-k^2 D_x,\\
\label{tm2v2}
&& \partial_{zx} E_x-\partial_{xx} E_z=k^2D_z,\\
\label{tmMagnet}
&& \partial_z H_y=ikc\epsilon_0D_x.
\end{eqnarray}
While for the TE-polarized modes, only $E_y$, $H_x$ and $H_z$ are non-zero
and the resulting equations are
\begin{eqnarray}
\label{te1}
&& \partial_{zz} E_y= - k^2 D_y,\\
&& \frac{k}{c\epsilon_0}H_x=i\partial_zE_y,~\frac{k}{c\epsilon_0}H_z=-i\partial_xE_y \label{te2}.
\end{eqnarray}
We only need to solve these equations for the electric field components in both the TE and TM cases.
The constitutive relation is taken as for isotropic materials
\begin{equation}
\vec {D}=\epsilon \vec {E} +
\frac{1}{2}\chi_3 [ |\vec{E}|^2\vec {E}
+ \frac12 (\vec{E}\cdot\vec{E})\vec{E}^*],
\label{eqD}
\end{equation}
where $\vec D$ is the displacement in SI units normalized to $\epsilon_0$.
The above expression for
$\vec D$ is an approximation for anisotropic semiconductors,
but its use is sufficient to demonstrate the reality
of the effects we are interested in and helps to improve the transparency of
our results and to simplify the complex numerical calculations. We also neglect
linear and nonlinear absorptions, which influence, but do not prevent observation
of soliton effects \cite{Ding2008,bullets}.

In our numerical approach, we do not force boundary conditions at the interfaces, but instead assume
that the linear permittivity $\epsilon$ and the nonlinear susceptibility $\chi_3$ change
continuously (but sharply) between their respective values for the material $s$ (silicon)
and material $g$ (silica glass).
We model the structure by taking
\begin{equation}
\epsilon(x)=\epsilon_{g}+\sum_{j}  (\epsilon_{s}-\epsilon_{g})
K_j(x),
\end{equation}
where $K_j(x)=\exp\{-[(x-x_j)/w]^{10}\}$ is the array of super-gaussian functions,
$j=0,\pm1,\pm 2\dots$, $x_j=j(s+w)$ is the position of the $j$th semiconductor layer,
$w$ is its width and $s$ is the side to side separation of the semiconductor layers (see Fig. 1),
$\epsilon=n^2$ and $n_g=1.44$, $n_s=3.48$.
$\chi_3$ is linked to the $n_2$ coefficient (measured
in $m^2/W$) and found in the tables as
$\chi_3={4\over 3}n_2\epsilon\epsilon_0c$ \cite{agrawal}. Thus it is convenient
to introduce  the function
\begin{equation}
\gamma(x)=\epsilon_0cn_2(x)\epsilon(x),
\end{equation}
 where
\begin{equation}
n_2(x)=n_{2,g}+\sum_{j}(n_{2,s}-n_{2,g})K_j(x),
\end{equation}
$n_{2,g}=2.5\cdot 10^{-20}$m$^2$/W, $n_{2,s}=4\cdot 10^{-18}$m$^2$/W .

\subsection{Equations for TM solitons}
We seek soliton solutions of Eqs.~(\ref{tm1v2}), (\ref{tm2v2}), (\ref{eqD})
in the form
\begin{equation}
E_x=f(x)e^{iqkz},~E_z=ig(x)e^{iqkz}.
\end{equation}
After some algebra, we find  that the real functions $f$ and $g$  obey
the system of the first order ordinary differential equations
\begin{eqnarray}
\label{tmfin1}
f^{\prime}&=&\frac{k\partial F /\partial g-q\left[
\epsilon^{\prime} + \gamma^{\prime} (f^2 + g^2/3)\right]f}
{q \left\{\epsilon+\gamma[3f^2+g^2/3]\right\}}\;,   \\
\label{tmfin2}
g^{\prime}&=&-\frac{k\partial F /\partial f}
{q \left\{\epsilon+\gamma[3f^2+g^2/3]\right\}},
\end{eqnarray}
where prime indicates first derivative in $x$ and the parameter $q$ measures the
relative change of the propagation constant with respect to its vacuum value $k$.
The function $F(f,g)$ is given by
\begin{eqnarray}
\nonumber
F&=&\frac{\gamma^2f^6}{2}+\frac{\gamma [4\epsilon - 3q^2]f^4}{4}+\frac{\epsilon [\epsilon-q^2]f^2}{2}
+\frac{\gamma q^2 g^4}{4}
\\
\label{integral}
&+&\frac{q^2\epsilon g^2}{2}
+\frac{\gamma f^2g^2}{3}\left[\gamma\left(f^2 + \frac{g^2}{6}\right)+
\epsilon-\frac{q^2}{2}\right]\;.
\end{eqnarray}
 $F$ becomes the first integral of Eqs.~(\ref{tmfin1})-(\ref{tmfin2}) in the case
 of a homogeneous medium (when $\epsilon$ and $\gamma$ are $x$ independent) \cite{CCP+2005}.

\subsection{Equation for TE solitons}
Soliton solutions of Eqs. (\ref{te1}), (\ref{eqD}) are sought in the form
\begin{equation}
E_y=u(x)e^{iqkz},
\end{equation}
which results in the familiar stationary nonlinear Schr\"odinger equation (NLSE)
\begin{equation}
u^{\prime\prime}+k^2(\epsilon-q^2)u+k^2\gamma u^3=0.
\end{equation}
In a case of the evanescently coupled semiconductor waveguides, the
above equation is readily transformed into a set of coupled mode
algebraic equations, giving familiar discrete soliton solutions
\cite{LSC+2008}. In the limit when $w$ and $s$ are much less then
$\lambda_{vac}$, one should expect the soliton profiles to be close
to the ones known from the NLSE with constant coefficients. Such
qualitative conclusions are, however, difficult to make simply by
looking at the equations for TM solitons. {\bf A} useful insight
into the difference between the TM and TE waves can be obtained when
considering linear modes of the structure.

\section{Linear modes, band structure and Brewster condition}
For $\gamma\equiv 0$
Eqs.~(\ref{tmfin1})-(\ref{integral}) for TM waves
are reduced to the linear eigenvalue problem
\begin{equation}
\label{linear}
q^2 k^2 f=\left(f^{\prime\prime}+\frac{\epsilon^{\prime}}{\epsilon}f^{\prime}
\right)+\left[
k^2\epsilon+\frac{\epsilon^{\prime\prime}}{\epsilon}-\left(\frac{\epsilon^{\prime}}{\epsilon}\right)^2
\right]f\;.
\end{equation}
It can be seen that the sharp jumps in $\epsilon$
can be compensated  only if the function
$f$ itself  changes sharply.
The jump in $f$ is determined by the refractive index contrast,
and follows from the continuity of $D_x$ at an  interface.

For TE waves the linear equation is
\begin{equation}
\label{linear_te}
k^2q^2u=u^{\prime\prime}+k^2\epsilon u.
\end{equation}
Therefore, jumps in $\epsilon$ only force the second derivative of $u$
to change accordingly, while $u$ itself stays smooth.

According to the Floquet-Bloch theorem, linear modes for TE and TM cases ($f$ or $u$)
can be represented in the form
$r(x)\exp(i k_B k x)$, where $r(x+w+s)=r(x)$
and $w+s$ is the period. All eigenvalues $q^2>0$ are parameterized
by the Bloch wavenumber $k_B$, $0\le |k_B|\le \pi/(k(w+s))$.
In the linear case, the problem is
tractable analytically with exact boundary conditions at the interfaces \cite{book_yeh,josaa}.
The discussions in  \cite{book_yeh,josaa} are focused on the band structure and miss
important for us features of the linear mode profiles, which  are
highlighted  below.

\begin{figure*}
\includegraphics[width=0.95\textwidth]{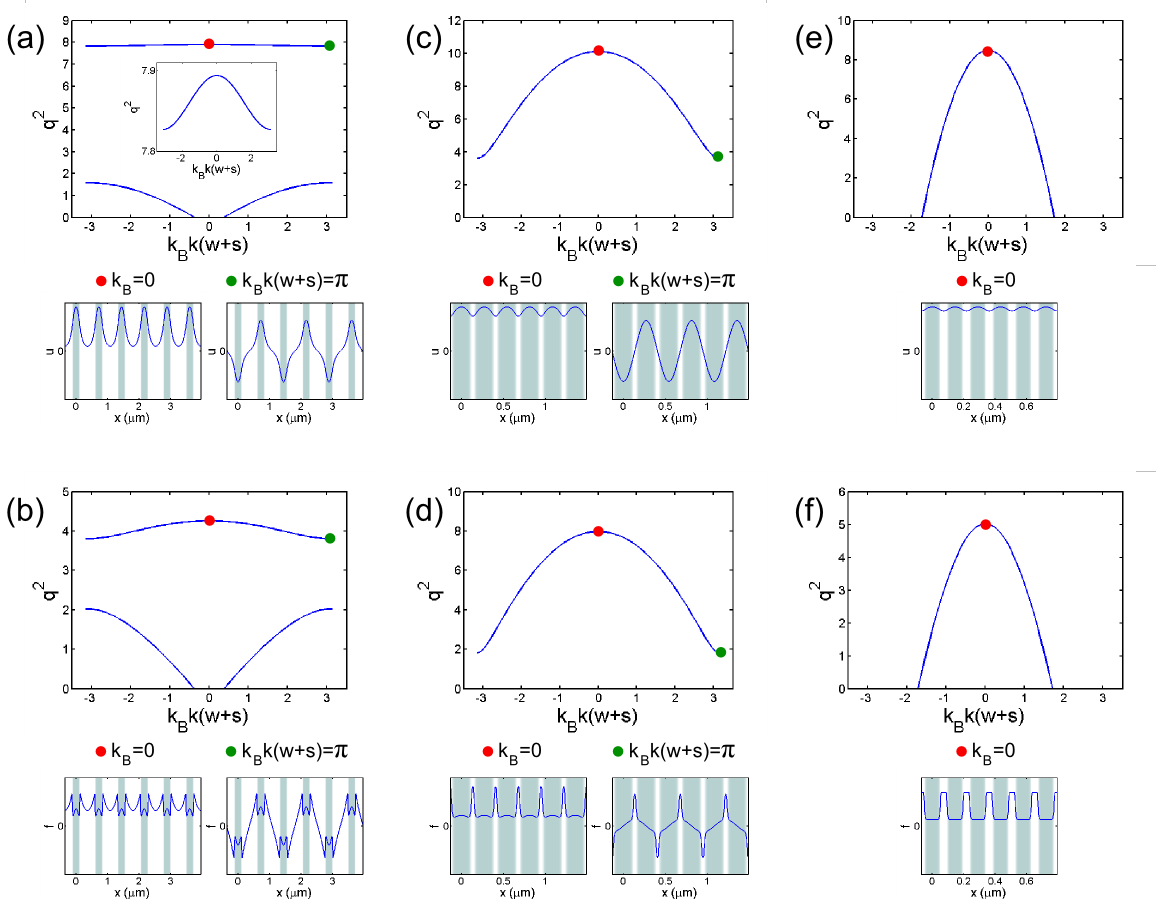}
\caption{(Color online) Spectra of linear TE (a,c,e) and TM (b,d,f) waves in different periodic structures:
(a),(b) $w=220$nm, $s=500$nm; (c),(d) $w=220$nm, $s=50$nm; (e), (f) $w=95$nm, $s=50$nm.
Small plots show the mode profiles at the centre and edge of the of the Brillouin zone from the top band.
The inset in (a) shows zoom of the top band.}
\label{fig2}
\end{figure*}

If we choose $w=220$nm and a sufficiently large separation  between the semiconductor
layers $s=500$nm, then
the  spectrum $q^2(k_B)$  for both TE and TM modes has two bands, see  Figs.~\ref{fig2}(a) and (b).
The $k_B=0$ and $k_B=\pi/(k(w+s))$ Bloch modes of the top band consist, respectively,
of the in-phase and out-of-phase modes of the individual waveguides, see insets in Figs.~\ref{fig2}(a) and (b).
The TE  Bloch modes can be well approximated using the well known  tight-binding approach \cite{LSC+2008},
when the Bloch mode is represented
by a superposition of the modes of the individual semiconductor layers. For TM modes, the field structure
is  dominated by jumps at the boundaries. However, the field overlap in the dielectric (silica)
layers still happens via exponentially decaying tails, and so the
tight binding approximation is justifiable here as well. In both cases, the tight binding model
will readily reproduce the almost sinusoidal profile
of $q^2(k_B)$ for the top band.
The  modes of the top band  with small $k_B$
experience normal (as in free space) diffraction, while those with $k_B$ near $\pi$ have
anomalous diffraction. This is in contrast to the system of coupled metallic slot waveguides, where
the situation is the opposite due to $\epsilon$ being negative in metals \cite{LBG+2007}.

As we reduce $s$, whilst keeping $w=220$nm, we find that for TM modes,  the gap below the
top band  shrinks and disappears at the  Brilluoin zone edges. The gap closure happens when
 $s=(\pi\sqrt{\epsilon_{s}+\epsilon_{g}}-\epsilon_{s}kw)/\epsilon_{g}/k\equiv s_0$.
This result is derived using the exact boundary conditions at the
interfaces and is reproduced well while numerically solving  Eq.
(\ref{linear}) ($s_0\approx 125$nm for $w=220$nm). The critical
value $s_0$ exists due to the  Brewster angle condition, which gives
zero reflection of the TM polarized wave from an interface. When the
Brewster condition is satisfied, then the
resonant Bragg scattering is canceled, the gap is closed and
 the periodic medium becomes  transparent \cite{book_yeh,josaa}.
For $s<s_0$ the gap in $q$ opens again, while  the lower  band continues to sink and soon disappears below
the $q^2=0$ line. The band structures of TM waves and Bloch modes in this regime
are shown in Fig.~\ref{fig2}(d).

Crucially,  at the instant when $s<s_0$ the geometry of the top band  TM modes
undergoes structural transformation. Specifically, the out-of-phase modes at the bottom
of the band now cross zero not inside the low index material, but inside the high index one,
cf. insets in Fig.~\ref{fig2}(b) and (d).
 This transformation signals a qualitative transition to the regime,
where it is appropriate to consider our periodic medium as an array of coupled slot waveguides
with sub-wavelength light localization inside the low index slots.
Indeed, the field intensity of the TM modes at the top
of the band now peaks
in-between the semiconductor layers and is strongly depressed inside them, see inset in Fig.~\ref{fig2}(b).

In contrast to the TM modes, the first band gap of the TE modes never shrinks to zero,
and the structural transformation of the  profiles of the top-band TE modes does not happen, cf.
Figs.~\ref{fig2}(a) and (c). The lower band of the TE spectrum sinks below the $q^2=0$ line
in a way similar to the TM case.

Further reduction of either $s$ or $w$ (or both of them simultaneously) leads
to the edges of the transmission band sinking below the $q^2=0$ cutoff for
both the TE and TM modes, so that the out-of-phase
modes corresponding to anomalous diffraction  gradually  disappear, see Figs.~\ref{fig2}(e) and (f).
The linear spectrum  in this regime qualitatively reproduces that of a
homogeneous medium, so that the description of the structure with an effective index approach becomes relevant.
The TE modes in this regime approach the ones of a homogeneous medium. E.g.,
the TE mode in the middle of the Brillouin
zone tends to a constant, see inset in Fig.~\ref{fig2}(e). At the same time the TM modes remain deeply
modulated, with pronounced jumps at interfaces, see inset in Fig.~\ref{fig2}(f).

The excitation of a periodic medium  with a narrow  beam
naturally results in  a diffractive spreading.
We can estimate the diffraction length $l_d$ of a beam with a radius $d$ as $l_d=kd^2/|\delta|$,
where the diffraction coefficient
\begin{equation}
\delta={d^2q\over dk_B^2}.
\end{equation}
 Provided the initial excitation has a flat phase across the beam, the
$\delta$ is calculated at $k_B=0$ of the highest band.
For $s=50$nm and $w=220$nm or $w=95$nm
the diffraction length of the $500$nm wide beam is approximately $3\mu$m
and is roughly the same for TE and TM waves. However, for $w=500$nm
the $l_d$ for TM modes is still around $3\mu$m, while for TE modes it is about an order
of magnitude more. We note that the overlap  of the TE modes through
the evanescent fields inside the dielectric layers is
smaller than the overlap of the TM modes. This is consistent with the difference
of the diffraction lengths.

Using the soliton concept we aim to demonstrate that preventing  this fast spreading and achieving
subwavelength localisation of light is possible.  Below we are establishing
existence and study stability of  solitons and, on this basis,
are making some assumptions about their mobility. The propagation studies, which will allow full exploitation of
soliton dynamics and functionality,  require development
of a different set of numerical tools and are postponed for the future.

\section{Numerical methods for finding soliton solutions and determining their stability}
Soliton solutions  of Eqs.~(\ref{tmfin1})-(\ref{integral}) discussed in the next chapter
have been  found using the shooting method with zero boundary conditions at infinity.
Since the nonlinearities of silica and silicon are positive, the soliton propagation constant $q$
has to be greater than the one for the linear waves, i.e.  $q>q_{lin}$. Here
$q_{lin}$ is the  propagation constant of the $k_B=0$
linear mode of the top band.

To characterize soliton solutions we use the  power density $P_z$
defined as the $x$ integrated $z$ component of the time averaged Poynting vector
$\left< \mathcal{\vec{S}} \right>$,
\begin{equation}
P_z=\int_{-\infty}^{\infty}  \left< \mathcal{S}_z\right>dx,~ \mathcal{\vec{S}}=\vec{\cal
E}\times\vec{\cal H}\;,
\end{equation}
where $\left< .. \right>$ denotes time averaging.
We plot $P_z$ as the function of the nonlinear phase shift  induced by a soliton.
The phase shift is defined as the difference between the total and
the maximal linear wavenumbers
\begin{equation}
\phi=k(q-q_{lin}).
\end{equation}

Substituting
$\vec{E}(x,z)=(\vec{E}_0(x)+ \vec{e}(x,z))\exp(iqk z)$,
$\vec{H}(x,z)=(\vec{H}_0(x)+ \vec{h}(x,z))\exp(iqk z)$ and
linearizing the Maxwell equations for small  $\vec{e},\vec{h}$ we
find that the latter obey:
\begin{eqnarray}
\label{evprob1}
i\partial_z \vec{a}&=&k\hat{A}\vec{a}+\hat{K}\vec{b}\;,\\
\label{evprob2}
k\hat{L}\vec{b}&=&\hat{M}\vec{a}\;.
\end{eqnarray}
Here $\vec E_0$ and $\vec H_0$ are the soliton solutions,
$\vec{a}=[e_x,e_y,h_x,h_y]^T$, $\vec{b}=[e_z,h_z]^T$,
\begin{equation}
\label{matrixA}
\hat{A}=
\left[
\begin{array}{cccc}
q & 0 & 0 & -1/(c\epsilon_0) \\
0 & q & 1/(c\epsilon_0) & 0 \\
0 & c\epsilon_0(\epsilon+\nu_{yy}) & q & 0 \\
-c\epsilon_0(\epsilon+\nu_{xx}) & 0 & 0 & q
\end{array}
\right],
\end{equation}

\begin{equation}
\label{matrixKandM}
\hat{K}=
\left[
\begin{array}{cc}
i\partial_x & 0  \\
0 & 0 \\
0 & i\partial_x \\
 kc\epsilon_0\nu_{xz} & 0
\end{array}
\right],
\hat{M}=
\left[
\begin{array}{cccc}
0 & i\partial_x & 0 & 0 \\
0 & 0 & 0 & i\partial_x
\end{array}
\right],
\end{equation}

\begin{equation}
\label{matrixL}
\hat{L}=
\left[
\begin{array}{cc}
0 & -1/(c\epsilon_0)  \\
c\epsilon_0(\epsilon+\nu_{zz}) & 0
\end{array}
\right],
\end{equation}
and
\begin{eqnarray}
\label{alpha}
&&\nu_{xz}=\frac{2\gamma}{3}\left[
E_{x0}E_{z0}^*+c.c.
\right],\\
&& \nu_{ii}=\frac{2\gamma}{3}\left[
|\vec E_0|^2+2E_{i0}E_{i0}^* \right],~i=x,y,z
\end{eqnarray}

By noting that $\hat{L}$ can always be inverted
and assuming $\vec{a}(x,z)=\vec{\alpha}(x)\exp(-ik\lambda z)$,
we arrive at the eigenvalue problem
\begin{equation}
\label{evfin}
k^2\lambda \vec{\alpha}=\left(k^2\hat{A}+\hat{K}\hat{L}^{-1}\hat{M}\right)\vec{\alpha}\;.
\end{equation}
We approximate the derivatives in the matrices $\hat{K}$ and $\hat{M}$ by
finite differences and are interested in the spatially localized
$\vec\alpha$ only. Then any  $\lambda$ found with $Im\lambda>0$
corresponds to an  unstable  perturbation exponentially growing with
the propagation coordinate $z$.

\section{Numerical results}
For TM and TE waves we looked for and found two  families of
soliton solutions, which  differ  by the position of their center of
symmetry. First, is the family of {\em on-site solitons}, where the
soliton center of symmetry is located in the middle of one of the
semiconductor layers. So that, by saying {\em site} we assume a
semiconductor layer. Second is the family of {\em off-site
solitons}, where the soliton center of symmetry is located between
the semiconductor layers, i.e. within a slot filled with silica glass.

\begin{figure}
\includegraphics[width=0.22\textwidth]{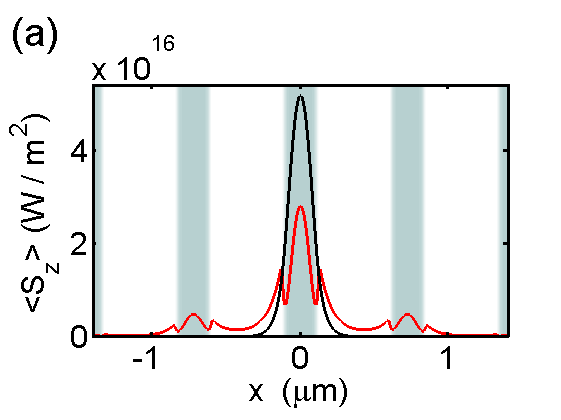}
\includegraphics[width=0.22\textwidth]{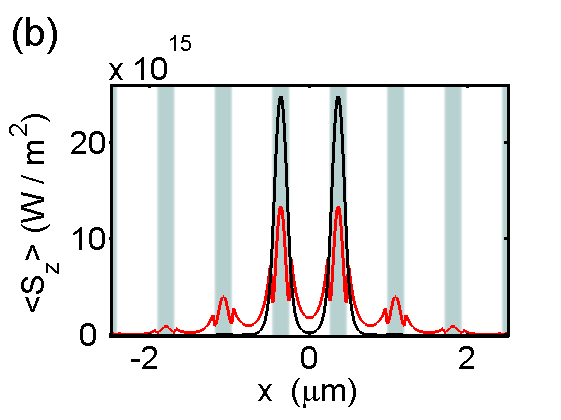}
\caption{(Color online) Intensity profiles of TE (black lines) and TM (red/grey lines)
solitons: (a) on-site; (b) off-site. The power density for all solitons is fixed to $P_z=10$ GW/m,
see the horizontal  lines and markers A and B in Fig.~\ref{fig4}.
Grayscale background illustrates the underlying periodic structure: $w=220$nm, $s=500$nm.}
\label{fig3}
\end{figure}

\begin{figure}
\includegraphics[width=0.22\textwidth]{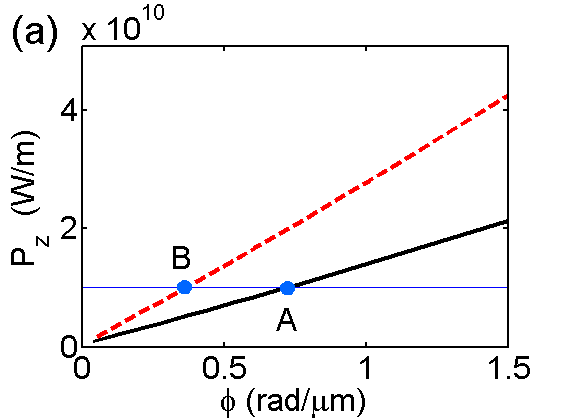}
\includegraphics[width=0.22\textwidth]{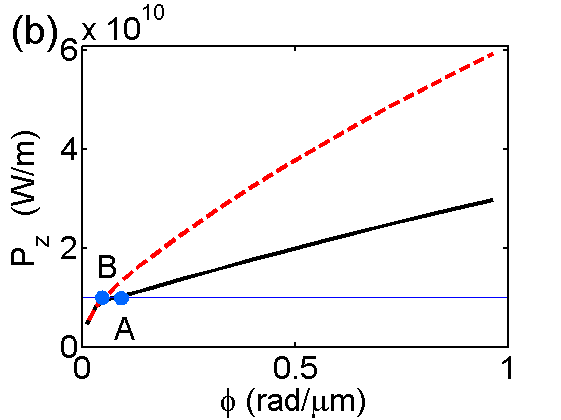}
\caption{(Color online) Power density as a function of the phase shift for TE (a) and
TM (b) solitons. Black (red/grey) lines correspond to the on-site (off-site)
soliton. Solid (dashed) lines indicate stable (unstable)
soliton branches. Structure parameters: $w=220$nm, $s=500$nm. Markers A and B
correspond to the on-site and off-site solitons shown in Fig.~\ref{fig3}.}
\label{fig4}
\end{figure}

\begin{figure}
\includegraphics[width=0.22\textwidth]{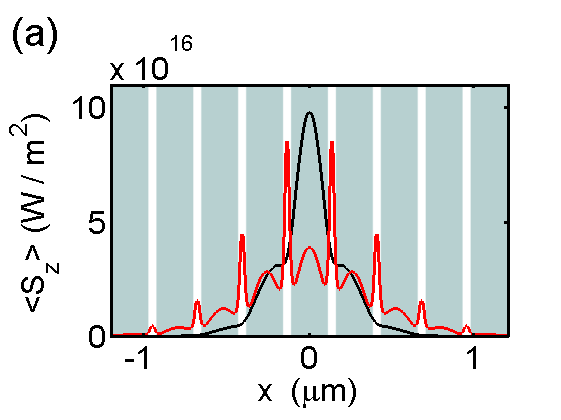}
\includegraphics[width=0.22\textwidth]{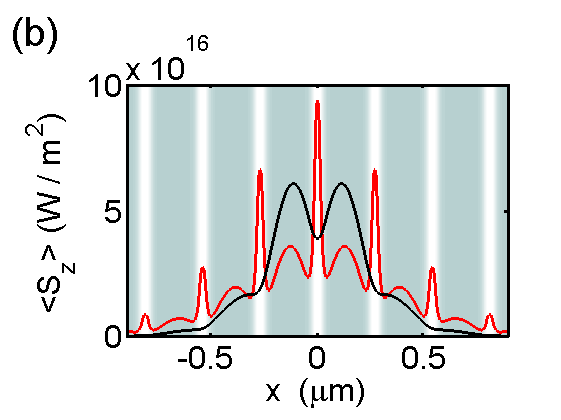}
\includegraphics[width=0.22\textwidth]{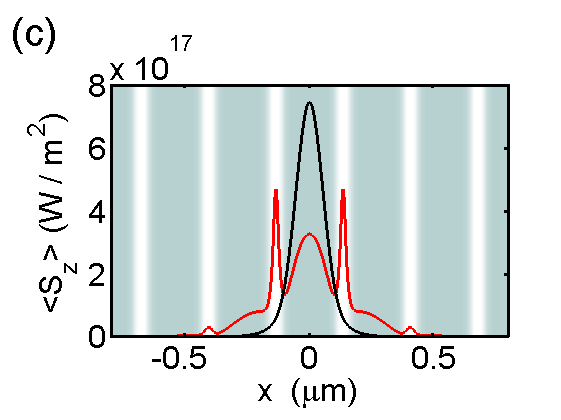}
\includegraphics[width=0.22\textwidth]{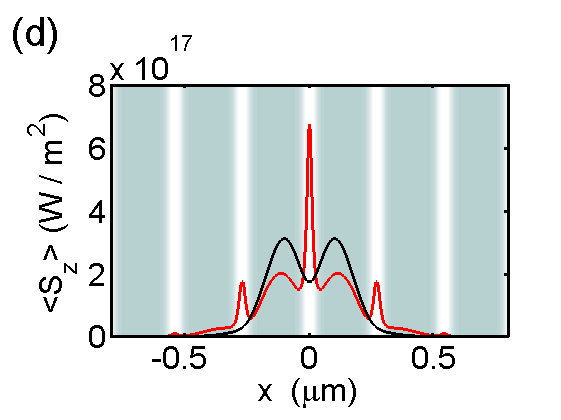}
\caption{(Color online) Intensity profiles of TE (black lines) and TM (red/grey lines)
solitons: (a,c) on-site; (b,d) off-site:  $w=220$nm, $s=50$nm.
The power density for  the solitons shown is fixed to: $P_z=33$ GW/m (a,b) and $P_z=110$ GW/m
(c,d), see the horizontal  lines and markers A,B,C,D in Fig.~\ref{fig6}.}
\label{fig5}
\end{figure}

\begin{figure}
\includegraphics[width=0.22\textwidth]{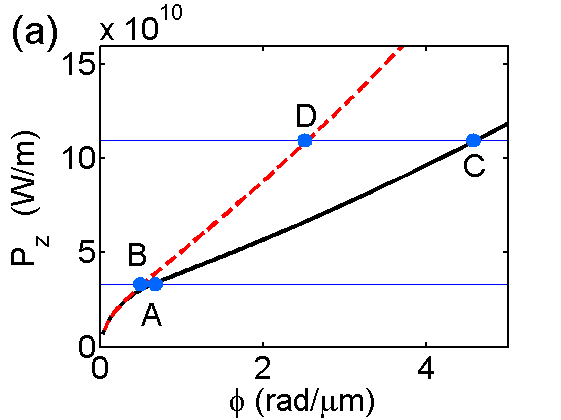}
\includegraphics[width=0.22\textwidth]{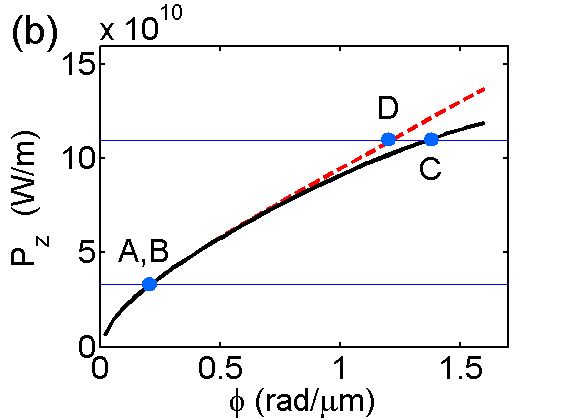}
\caption{(Color online)
Power density as a function of the phase shift for TE (a) and
TM (b) solitons. Black (red/grey) lines correspond to the on-site (off-site)
soliton. Solid (dashed) lines indicate stable (unstable)
soliton branches. Structure parameters: $w=220$nm, $s=50$nm.
Markers A-D correspond to the on-site and off-site solitons from
Fig.~\ref{fig5}(a)-(d), respectively.}
\label{fig6}
\end{figure}

\begin{figure}
\includegraphics[width=0.22\textwidth]{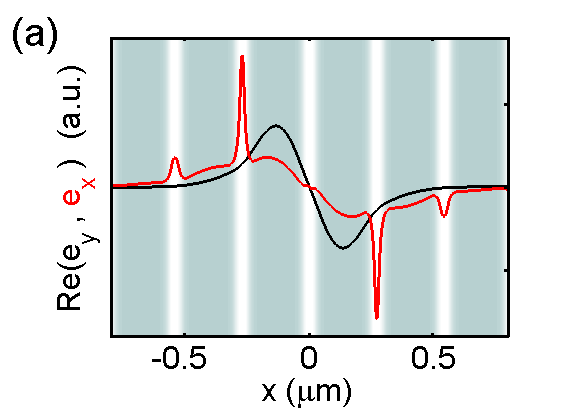}
\includegraphics[width=0.22\textwidth]{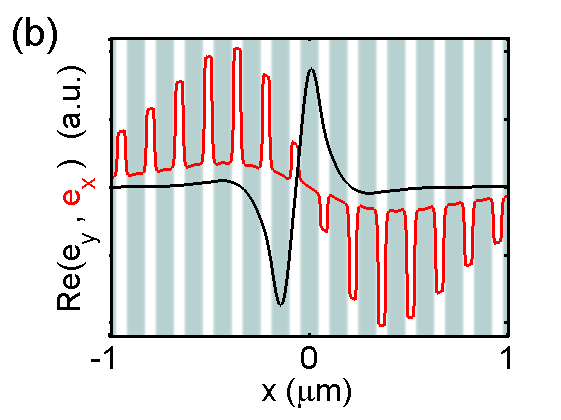}
\caption{(Color online) Unstable eigenmodes  for TE (black lines)
and TM (red/grey lines) solitons: (a) off-site TE and TM solitons from Fig.~\ref{fig5}(d);
(b) off-site TE soliton from Fig.~\ref{fig7}(d) and on-site TM soliton from Fig.~\ref{fig7}(c).}
\label{figUnst}
\end{figure}

\begin{figure}
\includegraphics[width=0.22\textwidth]{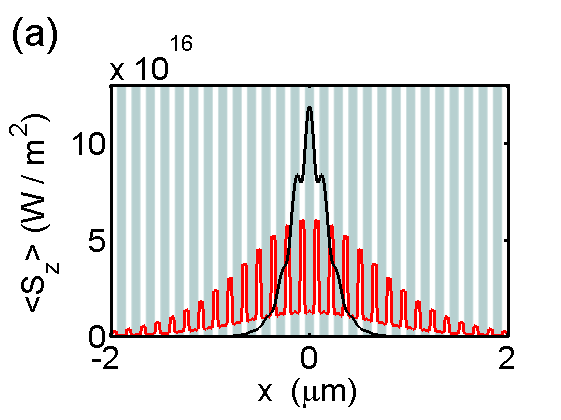}
\includegraphics[width=0.22\textwidth]{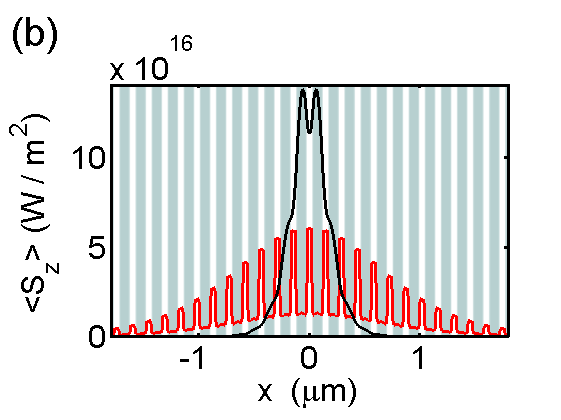}
\includegraphics[width=0.22\textwidth]{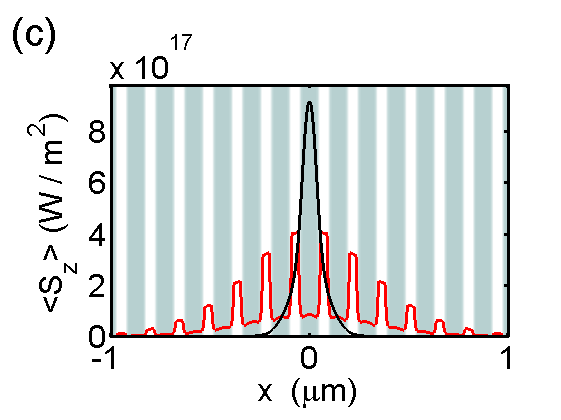}
\includegraphics[width=0.22\textwidth]{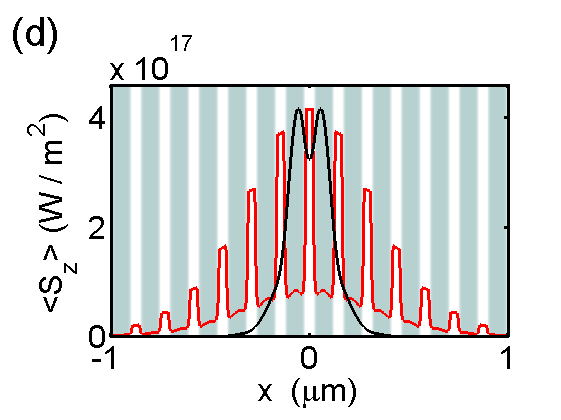}
\caption{(Color online)
Intensity profiles of TE (black lines) and TM (red/grey lines)
solitons: (a,c) on-site; (b,d) off-site:  $w=95$nm, $s=50$nm.
Power density for  the solitons shown is fixed to: $P_z=50$ GW/m (top row) and $P_z=110$ GW/m
(bottom row), see the horizontal  lines and markers A,B,C,D in Fig.~\ref{fig8}.
}
\label{fig7}
\end{figure}

\begin{figure}
\includegraphics[width=0.22\textwidth]{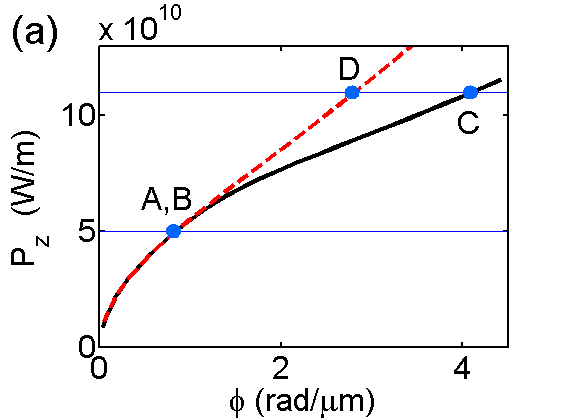}
\includegraphics[width=0.22\textwidth]{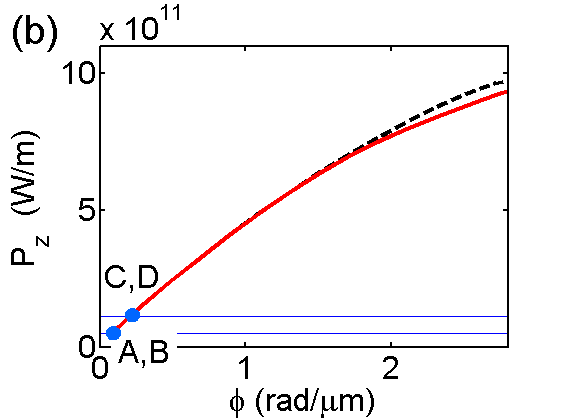}
\caption{(Color online)
Power density as a function of the phase shift for TE (a) and
TM (b) solitons. Black (red/grey) lines correspond to the on-site (off-site)
soliton. Solid (dashed) lines indicate stable (unstable)
soliton branches. Structure parameters: $w=95$nm, $s=50$nm.
Markers A-D correspond to the on-site and off-site solitons from
Fig.~\ref{fig7}(a)-(d), respectively.
}
\label{fig8}
\end{figure}

On-site and off-site soliton profiles for $w=220$nm and
$s_{0}<s=500$nm are shown in Fig.~\ref{fig3}.
One can see that for this relatively large separations, the
TE and TM solitons have a similar field structure with the light mostly concentrated
inside the semiconductor layers.
Such field profiles can be well approximated
by considering evanescently coupled semiconductor waveguides. The TM soliton is broader
for the same power, since the modulus of the corresponding diffraction coefficient
is by order of magnitude larger: $\delta_{TM}\simeq -0.32$ and $\delta_{TE}\simeq -0.04$.
The on-site solitons are stable in this case, while the off-site ones
are unstable.  Note, that the on-site and off-site solitons in waveguide arrays
described by the discrete nonlinear Sch\"odinger equation have the same
stability property \cite{LSC+2008,my_db_review}. Power $P_z$ as a function
of $\phi$ is shown in
Fig.~\ref{fig4}. The on-site and off-site solitons make the
same phase shift providing that the latter has a higher power. This is
true for both the TE and TM families.

As  $s$ approaches  and becomes less than $s_0$, while $w$ is kept fixed at
$220$nm, the TM solitons undergo a qualitative change
similar to the changes in the linear Bloch modes.  The
intensity peaks within the slots start to grow and prevail, so that
the on-site  TM solitons now have a two-peak structure,
see Fig.~\ref{fig5}(a,c), while the off-site ones have a single dominant peak,
see Fig.~\ref{fig5}(b,d). Note, that we are still in the regime, when the entire top band
of Bloch modes has $q^2>0$.

The power $P_z$ as the function of the phase
shift $\phi$ is shown in Figs.~\ref{fig6}(a) and (b) for the TE and TM solitons,
respectively. For both the TE and TM families, the on-site solitons still
give a larger phase shift at a given power than the off-site solitons.
The unstable eigenmodes of the
off-site solitons from Fig.~\ref{fig5}(d) are shown in Fig.~\ref{figUnst}(a).
In both the TE and TM cases,
the  instability is associated with the anti-symmetric eigenmode
(known as "depinning mode" in tight-binding models \cite{my_db_review}) and is
therefore expected to induce soliton motion across the structure, resulting in emission of
dispersive waves and gradual convergence to a stable on-site soliton.

We now reduce the width of the silicon layers to $w=95$nm and enter
the regime, when the linear diffraction law starts to approach the
limit of a homogeneous material (i.e. when the edges of the
top band sink below $q^2=0$ line and anomalous diffraction regions disappear,
see Figs.~\ref{fig2}(e),(f)). In this regime, nothing new
happens to the TE solitons and their shape feels only very little of
the material inhomogeneity, see Fig.~\ref{fig7}. At the same time, the
structure of the TM solitons is still very imhomogeneous with sharp
peaks within silica slots, see Fig.~\ref{fig7}. As a significant fraction of light intensity
in TM solitons is now concentrated inside slots with low $n_2$, TM solitons are noticeably broader
than TE solitons at the same level of power, despite the corresponding diffraction
coefficient being larger by modulus
for TE top band mode: $\delta_{TM}\simeq -0.3$, $\delta_{TE}\simeq -0.4$.
Powers as function of $\phi$ for this geometry are plotted in Fig.~\ref{fig8}. While nothing qualitatively changes
for TE solitons, on-site and off-site TM solitons exchange their roles: now they produce
the same phase shift provided the  former has higher power. Importantly, this
power exchange is accompanied by the exchange in stability
as well. The off-site soliton has become stable, while the on-site has got
the instability with respect to  the anti-symmetric
linear eigenmode, see Fig.~\ref{figUnst}(b). This result strongly suggests
that the role of an elementary waveguide (or {\em site})
should be reconsidered for TM solitons in nanostructured periodic medium with the linear spectrum
approaching the spectrum of the homogeneous material.
Indeed, since both linear and nonlinear modes show strong light confinement inside  slots,
it is  appropriate to consider a slot waveguide as an elementary structure. It means that if the
coupled mode approach is to be developed for such geometries in the future, it should use the
slot modes as the basis.

We also note that stabilization of the off-site and destabilization of the on-site  solitons
with respect to the antisymmetric depinning perturbations happens in the discrete
nonlinear Schr\"odinger equation if nonlinear coupling between the adjacent sites is included
\cite{OJE2003}.  Remarkably, in our case the coupling between the slot modes occurs
through overlap of the evanescent fields inside the silicon, which is 100 times more
nonlinear than the silica glass inside the slots.
Similarly to other systems \cite{OJE2003,my_db_review}, the stability
exchange between the on-site and off-site solitons is expected
to be accompanied  by an enhanced mobility of
strongly inhomogeneous TM solitons.
The possibility of such enhanced mobility is a problem opened for future investigation,
which can  reveal new possibilities for
all-optical signal steering and manipulation in sub-wavelength structures.
We have also checked for the existence of the cross-polarization
instabilities, but have not found any for the examples presented above.
This issue deserves further more detailed analysis.

\section{Summary}
Using the first principle nonlinear Maxwell equations we have developed
and applied numerical techniques  for finding  solitons in periodic
semiconductor-dielectric nanostructures and determining their stability.

When the separation between the high index (semiconductor)
layers in a periodic structure is reduced beyond the critical value (Brewster condition), the structure
of the antisymmetric  TM modes at the edges of the highest Brilluoin zone
undergoes qualitative transformation. Namely the zeros of these modes
now appear in the middle of the high index layers and not in the low index ones.
This signals transition to the regime when a periodic nanostructure
can be considered as an array of coupled slot waveguides for TM waves,
rather than an array of coupled high index
waveguides guiding by means of total internal reflection.
Reducing the interface separation further one
enters the regime where the diffraction law in a periodic nanostructure
becomes similar to the diffraction in a homogeneous medium.
The TE solitons in this quasihomogeneous limit  feel little or no  periodicity.
However, the transverse profiles of the TM solitons reflect the subwavelength periodic structure
and have dominant intensity peaks inside the low index slots.
In this regime the TM solitons with the center of symmetry in the middle
of a low index slot are stable while the ones centered on a high index
layer are unstable. This further facilitates the idea that the TM waves
in periodic nanostructures can be considered as modes of the
coupled slot waveguides and calls for developing of the corresponding analytic approaches.

\end{document}